\documentclass[prd,amsmath,preprintnumbers,superscriptaddress,nofootinbib,twocolumn]{revtex4}

\usepackage[dvips]{color,graphicx}
\usepackage{dcolumn}
\usepackage{bm}
\usepackage{amsmath}
\usepackage{amssymb}
\newcommand{\be}{\begin{equation}}

\newcommand{\ee}{\end{equation}}
\newcommand{\bea}{\begin{eqnarray}}
\newcommand{\eea}{\end{eqnarray}}

\begin{document}
\title{Probing the QCD vacuum with an abelian chromomagnetic field: A study within an effective model}
\author{L. Campanelli}\email{leonardo.campanelli@ba.infn.it}
\affiliation{I.N.F.N., Sezione di Bari, I-70126 Bari, Italia}\affiliation{Universit\`a degli Studi di Bari, I-70126
Bari, Italy}

\author{M. Ruggieri}\email{marco.ruggieri@ba.infn.it}
\affiliation{I.N.F.N., Sezione di Bari, I-70126 Bari, Italia}\affiliation{Universit\`a degli Studi di Bari, I-70126
Bari, Italy}
\date{\today}
\begin{abstract}
We study the response of the QCD vacuum to an external abelian chromomagnetic field in the framework of a non local
Nambu-Jona Lasinio model with the Polyakov loop. We use the Lattice results on the deconfinement temperature of the
pure gauge theory to compute the same quantity in the presence of dynamical quarks. We find a linear relationship
between the deconfinement temperature with quarks and the squared root of the applied field strength, $gH$, in
qualitative (and to some extent also quantitative) agreement with existing Lattice calculations. On the other hand, we
find a discrepancy on the approximate chiral symmetry restoration: while Lattice results suggest the deconfinement and
the chiral restoration remain linked even at non-zero value of $gH$, our results are consistent with a scenario in
which the two transitions are separated as $gH$ is increased.
\end{abstract}
 \maketitle \preprint{BA-TH/609-09}

\section{Introduction}
Quantum Chromodynamics (QCD) is nowadays regarded as the theory of strong interactions. The study of the QCD vacuum and
of the QCD phase diagram is one of the most intriguing research topic in modern physics. A deep understanding of the
vacuum would allow us to encompass the mechanisms on which color confinement and spontaneous chiral symmetry breaking
lie.

The major knowledge on the QCD phase diagram arises from Lattice QCD at finite temperature and zero (or small) baryon
chemical potential, see for example~\cite{Bazavov:2009zn,Aoki:2009sc,deForcrand:2008zi} and references therein for
recent results. At finite chemical potential Montecarlo simulations with three colors suffer the well known sign
problem~\cite{Ejiri:2004yw}. In order to circumvent this problem several approaches to Lattice calculations have been
suggested~\cite{D'Elia:2004at,D'Elia:2002gd,Allton:2005gk,Fodor:2001pe,Fodor:2007vv,de Forcrand:2003hx,D'Elia:2007ke}.
Beside them, the Nambu-Jona Lasinio model~\cite{Nambu:1961tp,revNJL} is a very popular tool for making predictions on
the QCD phase diagram in regions unaccessible by Lattice QCD.

Spontaneous chiral symmetry breaking is characterized in terms of the non vanishing and non perturbative chiral
condensate $\langle\bar{\psi}\psi\rangle$, which is an order parameter in the massless quark case. On the other hand,
the color confinement is well defined in terms of an order parameter only in the pure gauge theory. In this case it is
well known that the Polyakov loop $L$ can be regarded as the order parameter for the confinement-deconfinement
transition~\cite{Polyakovetal}. As a matter of fact, $L$ is a gauge singlet field that transforms non trivially under a
$Z_3$ transformation. Thus in the phase with $\langle L \rangle \neq 0$ the $Z_3$ symmetry is spontaneously broken,
while it is not broken if $\langle L \rangle = 0$. Since $\langle L \rangle = 0$ in the confined phase and $\langle L
\rangle \neq 0$ in the deconfined system, one argues that deconfinement-confinement transition can be described as a
restoration of the spontaneously broken $Z_3$ symmetry. Dynamical quarks break the $Z_3$ symmetry, thus the Polyakov
loop in presence of dynamical quarks is no longer a good order parameter for deconfinement. Nevertheless lattice
measurements show the existence of a $Z_3$ crossover in the range of temperature in which both chiral condensate and
particle susceptibility have a crossover too, denoting that the Polyakov loop is still a good indicator of the
deconfinement transition.

In this paper, we use an external abelian chromomagnetic field to probe the vacuum structure of QCD.  This is a trick
that allows us to learn something more about the QCD vacuum, by studying its response to external perturbations. This
approach has revealed to be successful in Lattice QCD calculations~\cite{Cea:2002wx,Cea:2005td,Cea:2007yv}. One of the
main results obtained is that the deconfinement temperature at zero baryon density decreases as the strength of the
applied field $gH$ is increased. In more detail, in the case of the pure glue theory, Lattice data on the deconfinement
temperature $T_c$ as a function of $\sqrt{gH}$ are well interpolated by a linear fit for $gH$ lower than a critical
value $gH_c$, and consistent with zero for $gH>gH_c$. In presence of dynamical quarks, Lattice data are consistent with
a linear behavior of $T_c(gH)$ up to $gH$ of the order of 1 GeV$^2$, while at the moment there are not data for larger
values of $gH$. Nevertheless the existing data are once again well interpolated by a linear fit. An extrapolation of
the fit to larger values of the applied field strength implies that there exists a critical value of the applied field,
$gH_c$, such that the deconfinement temperature is zero, and for $gH>gH_c$ the system is deconfined at any temperature.
Moreover, the Lattice results lead to the conclusion that the lowering of the deconfinement temperature is a
peculiarity of the non abelian theories~\cite{Cea:2005td}. A possible explanation of these facts, as suggested by the
authors in Refs.~\cite{Cea:2002wx,Cea:2005td,Cea:2007yv}, is that the QCD vacuum behaves as a relativistic color
superconductor, whose color superconductivity is destroyed by a strong enough external magnetic field.

In our calculations, we use the Nambu-Jona Lasinio model with the Polyakov loop (PNJL model in what
follows)~\cite{Meisinger:1995ih,Blaschke:2007np,Contrera:2007wu,Ciminale:2007ei,Fu:2007xc,Ciminale:2007sr,Hansen:2006ee,Sakai:2008py,Megias:2006bn,Fukushima:2003fw,Fukushima:2008wg,Abuki:2008ht,Abuki:2008nm,Abuki:2008tx,Roessner:2006xn,Ratti:2005jh,Ratti:2007jf,Hell:2008cc,Sasaki:2006ww,Schaefer:2007pw}
as an effective model of QCD. The PNJL model has been widely used in the literature to depict several aspects of the
QCD phase diagram. In the PNJL model, one has not dynamical gluons and the thermodynamics of the Polyakov loop $L$ is
driven by a temperature dependent effective potential ${\cal U}$ for $L$, the latter being coupled minimally to quarks
via the QCD covariant derivative. Instead to use the usual hard cutoff regularization scheme, we consider here a
non-local
interaction~\cite{Schmidt:1994di,Bowler:1994ir,Blaschke:2000gd,GomezDumm:2005hy,Aguilera:2006cj,Grigorian:2006qe,GomezDumm:2006vz}.
We take an instantaneous, rotationally and translationally invariant interaction, leaving the more general case of a
non instantaneous interaction to a future study. The non local interaction is modelled in momentum space by a form
factor which depends only on the magnitude of the $3-$momentum and on a mass scale $\Lambda$. This approach has the
advantage that it does not artificially cutoff large momenta, thus making the model suitable for field strengths $gH$
larger than the natural momentum scale $\Lambda$ of the form factor.

Since the PNJL model does not describe dynamical gluons it cannot be used to predict the response of the pure gauge
system to an external field. Therefore in this paper we use the Lattice QCD result on the dependence of the
deconfinement temperature of the pure gauge theory on $gH$ as an input. This dependence has the
form~\cite{Cea:2002wx,Cea:2005td,Cea:2007yv}
\begin{equation}
\frac{T_c(gH)}{T_c} = 1-\alpha\frac{\sqrt{gH}}{T_c}~, \label{eq:Intro1}
\end{equation}
where $T_c$ is the deconfinement temperature at zero field and $\alpha$ is a dimensionless constant. However we are
able to compute the effect of dynamical quarks on deconfinement, and compare them with Lattice QCD results. Remarkably
enough, our findings on deconfinement temperature with quarks are in agreement with Lattice QCD calculations. In
particular, we will show that the linear dependence depicted in Eq.~\eqref{eq:Intro1} is still valid in presence of
quarks, as it is observed in Lattice measurements. Even if the larger part of our results are obtained in the physical
limit in which $m_\pi = 135$ MeV and for a specific choice of the form factor, we have checked the stability of our
results by changing the analytical expression of the form factor in the chiral limit (finite quark masses are not
expected to play a relevant role in this context).

The plan of the paper is as follows. In section II, we review the PNJL model with non local instantaneous interaction,
and discuss the quark spectrum in the presence of an abelian external chromomagnetic field. In section III, we present
our results on chiral and deconfinement transitions, as well as the phase diagram of the model in the $gH-T$ plane and
the equation of state of the PNJL matter in the external field. Finally, in section IV we draw our conclusions.

\section{The PNJL model in external abelian chromomagnetic field}
Our investigation concerns the response of the PNJL vacuum to an external abelian chromomagnetic field specified by the
gauge potential
\begin{equation}
A_{a\mu}= Hx\delta_{a3}\delta_{\mu_2}~,~~~~~H>0~,\label{eq:H}
\end{equation}
corresponding to a field along the positive $z$-axes. The quark Lagrangian density is given
by~\cite{Fukushima:2003fw,Abuki:2008tx,Abuki:2008nm}
\begin{equation}
{\cal L}= \bar\psi\left(i\gamma_\mu D^\mu -m\right)\psi + {\cal L}_{4} - {\cal U}[L,\bar L,T]~. \label{eq:LagrP}
\end{equation}
In the above equation we have omitted the pure gauge contribution $\propto\bm H^2$ since it is a constant and thus not
relevant for the dynamics; $\psi$ is the quark field with Dirac, color and flavor indices (implicitly summed). $m$
corresponds to the bare quark mass matrix: we work with two flavors and we assume from the very beginning $m_u = m_d$.
The covariant derivative is defined as $D_\mu =
\partial_\mu -i A_\mu$ with $A_\mu = g A_\mu^a T_a$ and $T_a$, $a=1,\dots,8$ being the $SU(3)$ color generators with the normalization condition
$\text{Tr}[T_a,T_b]=\delta_{ab}$.

In Eq.~\eqref{eq:LagrP} $L$, $\bar L$ correspond to the normalized traced Polyakov loop and its hermitian conjugate
respectively, $L=\text{Tr}W/N_c$, $\bar L=\text{Tr}W^\dagger/N_c$, with
\begin{equation}
W={\cal P}\exp\left(i\int_0^\beta A_4 d\tau\right)=\exp\left(i \beta A_4\right)~,~~~~~A_4=iA_0~,
\end{equation}
and $\beta=1/T$. In the above equation we have implicitly assumed that the thermodynamics of the Polyakov loop is
described in terms of a constant and homogeneous background field $A_0$. Even if this choice leads to interesting
agreement of the PNJL model with lattice calculations, more sophisticated models with inhomogeneous background field
have been investigated, see for example~\cite{Megias:2006bn}.

\begin{widetext}
The term ${\cal U}[L,\bar L,T]$ in Eq.~\eqref{eq:LagrP} is the effective potential for the traced Polyakov loop; it is
built by hand in order to reproduce the pure glue lattice data of
QCD~\cite{Fukushima:2003fw,Ratti:2005jh,Roessner:2006xn,Ghosh:2007wy,Fukushima:2008wg}. In this paper, we work in the
Polyakov gauge in which
\begin{equation}
L = \frac{1}{3}\text{Tr}\left[e^{i \beta (\lambda_3 \ell_3 + i \lambda_8 \ell_8})\right]~,
\end{equation}
with $\ell_3$, $\ell_8$ real parameters. Since we are interested to the case of zero baryon chemical potential we set
$\langle L\rangle = \langle\bar L\rangle$ from now on. This choice implies $\ell_8 = 0$ and thus we are left with only
one parameter, $\ell_3\equiv\ell$. Moreover, we adopt the following logarithmic form~\cite{Roessner:2006xn},
\begin{equation}
{\cal U}[L,\bar L,T] = T^4\left[-\frac{b_2(T)}{2}\bar LL + b(T)\log\left[1-6\bar LL + 4(\bar L^3 + L^3) -3(\bar
LL)^2\right]\right]~,\label{eq:Poly}
\end{equation}
with
\begin{equation}
b_2(T) = a_0 + a_1 \left(\frac{\bar T_0}{T}\right) + a_2 \left(\frac{\bar T_0}{T}\right)^2~,~~~~~b(T) =
b_3\left(\frac{\bar T_0}{T}\right)^3~.\label{eq:lp}
\end{equation}
Numerical values of the coefficients are as follows~\cite{Roessner:2006xn}:
\begin{equation}
a_0=3.51~,~~~a_1 = -2.47~,~~~a_2 = 15.2~,~~~b_3=-1.75~.
\end{equation}
\end{widetext}The parameter $\bar{T}_0$ in Eq.~\eqref{eq:Poly} sets the deconfinement scale in the pure gauge theory.
Lattice results show that the deconfinement temperature of the pure glue system is a linear and decreasing function of
$\sqrt{gH}$~\cite{Cea:2002wx,Cea:2005td}, at least for small values of $\sqrt{gH}/T_c$, with $T_c$ the deconfinement
temperature at zero field. For $\sqrt{gH}$ larger than a critical value of the order of 1 GeV, the critical temperature
of the pure glue system as computed in Refs.~\cite{Cea:2002wx,Cea:2005td} are consistent with zero. Inspired by these
results, we make the following ansatz for $T_0$ in Eq.~\eqref{eq:Poly}:
\begin{eqnarray}
\bar{T}_0 &=& 270~\text{MeV}\times\theta\left(1- \frac{\sqrt{gH}}{\sqrt{gH_c}}\right)~,\label{eq:T0mM}
\end{eqnarray}
with $\theta$ denoting the unit step function, and we take $\sqrt{gH_c} = 1.2$ GeV according to the estimates
of~\cite{Cea:2002wx,Cea:2005td}.

Finally, ${\cal L}_4$ in Eq.~\eqref{eq:LagrP} represents the lagrangian density for the four fermion interaction. If we
define $S_4 = \int d^4 x {\cal L}_4$ as the interaction action
then~\cite{Sasaki:2006ww,Schmidt:1994di,Bowler:1994ir,Blaschke:2000gd,GomezDumm:2005hy,Aguilera:2006cj,Grigorian:2006qe}
\begin{equation}
S_{4} = G \int d^4 x~\left[(\bar q(x) q(x))^2 + (\bar q(x) i\gamma_5\bm\tau q(x))^2\right]~,\label{eq:1}
\end{equation}
with the dressed quark field defined as
\begin{equation}
q(x)= \int d^4 y~F(x-y) \psi(y)~,\label{eq:dress}
\end{equation}
where $F(r)$ is a form factor whose Fourier transform $f(p)$ satisfies the constraint $f(p)\rightarrow 0$ for
$p\rightarrow\infty$, $p$ being the 3-momentum. In this paper, we use the Lorentzian form factor
\begin{equation}
f(p)=\frac{1}{\sqrt{1+(p^2/\Lambda^2)^{N}}}~.\label{eq:f1}
\end{equation}
In the above equation $N=10$, $p=|\bm p|$ and $\Lambda = 684.2$ MeV. Moreover we use $m=4.46$ MeV and
$G=2.33/\Lambda^2$~\cite{Sasaki:2006ww}. By these numerical values one has $f_\pi = 92.3$ MeV and the pion mass $m_\pi
= 135$ MeV, as well as the zero momentum quark mass $M_u = 335$ MeV. The advantage of using a form factor in momentum
space is that it does not lead to the introduction of an artificial momentum cutoff as in the usual NJL model
calculations.

In order to study chiral symmetry breaking we assume that in the ground state
\begin{equation}
\sigma = G\langle \bar q(x)q(x)\rangle\neq0~,\label{eq:condensates}
\end{equation}
where a summation over flavor and color is understood.  In what follows we consider the system at finite temperature
$T$ in the volume $V$. This implies that the space-time integral is $\int d^4x = \int_{0}^\beta d\tau \int d^3\bm x$
with $\beta=1/T$. In order to define the thermodynamical potential at $H\neq0$, we observe that at $H=0$ the momentum
space mean field PNJL action reads~\cite{Abuki:2008nm}
\begin{eqnarray}
S&=&\int\frac{d^4p}{(2\pi)^4} \left[\bar\psi\left(\gamma_\mu p^\mu
-\gamma_\mu A^\mu \right)\psi\right] \nonumber\\
&&+\int\frac{d^4p}{(2\pi)^4} f(p)^2\left[2\sigma~\bar\psi(p) \psi(p)
\right] \nonumber\\
&&~~~- \beta V \frac{\sigma^2}{G} - \beta V {\cal U}[L,\bar L,T]~, \label{eq:LagrMFms}
\end{eqnarray}
with $V$ denoting the quantization volume and $A_\mu = g A_\mu^a T_a$. We introduce the mean field, momentum dependent,
constituent quark mass $M(p)$:
\begin{equation}
M(p) \equiv m-2\sigma f^2(p) \equiv m + \Sigma(p)~,\label{eq:mass}
\end{equation}
where $\Sigma$ denotes the momentum dependent proper quark self energy.

In order to compute the effective potential for $\sigma$ and $L$ we need to know the quark spectrum. The computation of
the latter in the case of non-local interaction is a non-trivial task, even in the mean field approximation: As a
matter of fact, the Dirac equation for the quarks in the external field is an integro-differential whose solution is
beyond the scope of the present study. For this reason, in this paper we will make a simple ansatz for the quark
spectrum of the non-local theory, arguing its specific form from that of the local NJL model. We recover the local NJL
model by taking $N\rightarrow\infty$ in Eq.~\eqref{eq:f1}. Hence, the larger the value $N$, the better is our
assumption.

\begin{widetext}
The thermodynamical potential $\Omega$ per unit volume at $H=0$ can be obtained by integration over the fermion fields in
the partition function of the model:
\begin{eqnarray}
\Omega &=&  {\cal U}[L,\bar L,T] + \frac{\sigma^2}{G} - T\sum_n\int \frac{d^3{\bm
p}}{(2\pi)^3}~\text{Tr}~\text{log}\frac{S^{-1}(i\omega_n,{\bm p})}{T}~,
\end{eqnarray}
where the sum is over fermion Matsubara frequencies $\omega_n = \pi T(2n+1)$, and the trace is over Dirac, flavor and
color indices. The inverse quark propagator is defined as
\begin{equation}
S^{-1}(i\omega_n,{\bm p})= \left[(i\omega_n+iA_4)\gamma_0 -{\bm\gamma}\cdot{\bm p} -M(p)\right] \otimes{\bm 1}_f~.
\label{eq:po}
\end{equation}
Performing the trace and the sum over Matsubara frequencies the effective potential for $L$, $\sigma$ at $H=0$ reads
\begin{eqnarray}
\Omega &=& {\cal
U}[L,\bar L,T] + \frac{\sigma^2}{G}  -2 N_c N_f \int\!\frac{d^3\bm p}{(2\pi)^3}E\nonumber \\
&& -4 N_f T\int\! \frac{d^3{\bm p}}{(2\pi)^3}~\text{log}\left[1+3 L e^{-\beta E} + 3L e^{-2\beta E} + e^{-3\beta E}
\right]~,\label{eq:O1h}
\end{eqnarray}
where $E=\sqrt{p^2+M(p)^2}$. \end{widetext}

The quark spectrum in absence of external field is given by
\begin{eqnarray}
\varepsilon_r^s &=& \pm i\ell \pm \sqrt{p^2+ M(p)^2}~,\label{eq:Epm10}\\
\varepsilon_g^s &=& \mp i\ell \pm \sqrt{p^2 + M(p)^2}~,\label{eq:Epm20}\\
\varepsilon_b^s &=& \sqrt{p^2 + M(p)^2}~,\label{eq:Epm30}
\end{eqnarray}
where the upper (lower) sign corresponds to particles (antiparticles).

In presence of the external field~\eqref{eq:H} the red and green quark spectra is modified. If $S_4=0$ then it is well
known that the effect of the applied chromomagnetic field on the quark spectra is both diamagnetic and paramagnetic.
The former corresponds to the quantization of the quark motion in the plane orthogonal to the field direction. The
latter differentiates the spin up and spin down motion by virtue of the usual term $\sim\bm\sigma\cdot\bm H$ in the
interaction hamiltonian.

If $S_4\neq0$ but the interaction kernel is local [which corresponds to the choice $F(x-u)=\delta^4(x-u)$ in
Eq.~\eqref{eq:dress}], then the quark spectra in the case $H\neq0$ is
\begin{equation}
E_r^s = \sqrt{p^2_z + p_{\perp,s}^2 + M^2}~,~~~~~E_b^s = \sqrt{p^2 + M^2}~,\label{eq:EpmLOCAL}
\end{equation}
where
\begin{equation}
p_{\perp,s}^2 = gH\left(n+\frac{1}{2}+\frac{s}{2}\right)~,~~~~~n=0,1,2,\dots
\end{equation}
and $M$ denotes the constituent quark mass. In the case of the non-local interaction, we {\em assume} that the quark
dispersion laws for red and green quarks are given by
\begin{equation}
E_r^s = \sqrt{p^2_z + p_{\perp,s}^2 + M(p^2_z + p_{\perp,s}^2)^2}~,\label{eq:Epm}
\end{equation}
while the spectrum of the blue quarks, which are not coupled to the external field, is still given by
Eq.~\eqref{eq:Epm30}. The assumption in Eq.~\eqref{eq:Epm} is partly justified by the observation that $S_4$ is
translationally and rotationally invariant, and it does not contain derivative couplings, and at the mean field level,
it simply amounts to the replacement of a constant constituent quark mass with a momentum dependent one, see
Eq.~\eqref{eq:Epm10}. The assumption~\eqref{eq:Epm} leads to the correct spectrum in the case $S_4$ can be treated as a
perturbation, that is in the case of large $gH$ with respect to $\sigma$. This assumption has been used several times
in the literature for studying chiral symmetry breaking in a strong magnetic field~\cite{Miransky:2002rp,Kabat:2002er}.

In Equation~\eqref{eq:Epm} the subscripts $r$, $b$ denote the quark color; $s = \pm1$ distinguishes the quark spin
projection along the applied field; finally $n$ labels the Landau level. We have not explicitly written the energy of
the green quarks since the former depends only on the absolute value of the charge, thus $E_g = E_r$.  Notice that we
have introduced a spin index also for blue quarks: this is done for notational convenience. We have explicitly written
the dependence on $p_\perp$ in Eq.~\eqref{eq:Epm}; in the following we will write $M(p)$ for simplicity, leaving
understood the dependence on the true quark momentum $p_z^2 + p_\perp^2$. Keeping into account that red and green
quarks have respectively $T_3$ color charges given by $\pm 1/2$, the phase space integral is
\begin{equation}
\int\!\frac{d^3{\bm P}}{(2\pi)^3} =\frac{gH}{4\pi}\sum_{n=0}^\infty \int_{-\infty}^{+\infty}\!\frac{dp_z}{2\pi}~,
\label{eq:psiZ}
\end{equation}
for red and green quarks, and
\begin{equation}
\int\!\frac{d^3{\bm p}}{(2\pi)^3} =\frac{4\pi}{8\pi^3}\int_0^{+\infty}p^2 dp~,\label{eq:psiZZ}
\end{equation}
for blue quarks. Taking into account that in the Polyakov gauge the effect of the Polyakov loop is merely a shift of
the quark chemical potentials to a complex value we can write the quark poles as
\begin{eqnarray}
\varepsilon_r^s &=& \pm i\ell \pm \sqrt{p^2_z + p_{\perp,s}^2 + M(p)^2}~,\label{eq:Epm1}\\
\varepsilon_g^s &=& \mp i\ell \pm \sqrt{p^2_z + p_{\perp,s}^2 + M(p)^2}~,\label{eq:Epm2}\\
\varepsilon_b^s &=& \sqrt{p^2 + M(p)^2}~,\label{eq:Epm3}
\end{eqnarray}
where the upper (lower) sign corresponds to particles (antiparticles). \begin{widetext}The thermodynamical potential at
$H\neq0$ can be obtained from the expression at $H=0$, see Eq.~\eqref{eq:O1h}, by replacing the quark energies and the
phase space integral as follows:
\begin{eqnarray}
\Omega &=& C + {\cal
U}[L,\bar L,T] + \frac{\sigma^2}{G}  -N_f \sum_{s=\pm1}\left[\int\!\frac{d^3\bm P}{(2\pi)^3}2E_r^s + \int\!\frac{d^3\bm p}{(2\pi)^3}E_b^s \right]\nonumber \\
&& -2 N_f \sum_{s=\pm1}T\int\! \frac{d^3{\bm P}}{(2\pi)^3}~\text{log}\left[1+3 L e^{-\beta E_r^s} + 3L
e^{-2\beta E_r^s} + e^{-3\beta E_r^s}   \right]~\nonumber\\
&&-2 N_f \sum_{s=\pm1}T\int\! \frac{d^3{\bm p}}{(2\pi)^3}~\text{log}\left[1+e^{-\beta E_b^s}   \right] +
2 N_f \sum_{s=\pm1}T\int\! \frac{d^3{\bm P}}{(2\pi)^3}~\text{log}\left[1+e^{-\beta E_b^s}   \right]~.~\nonumber\\
\label{eq:O1}
\end{eqnarray}\end{widetext}
In Eq.~\eqref{eq:O1} $C$ denotes an irrelevant additive constant that makes the thermodynamic potential equal to zero
at $T=0$, $\sigma=0$ for each value of $gH$. This subtraction of the perturbative part of the thermodynamical potential
makes the latter finite at every temperature. Equation~\eqref{eq:O1} is in agreement with that of
Ref.~\cite{Ebert:2006uh} in the limit of the pure NJL model, the latter being recovered in the limit $L\rightarrow 1$
after subtraction of ${\cal U}(L,\bar L, T)$.

The last line of Eq.~\eqref{eq:O1} corresponds to the thermal blue quark contribution {\em subtracted} of a fictitious
contribution of the quarks of the same color in the external field. The same contribution is {\em added} to the second
line. This procedure is applied for the phase space integral is different for red and green quarks on one hand and blue
quarks on the other hand, thus their contribution to the free energy cannot be collected under an unique integral sign.
This procedure has the advantage that it allows us to write the dependence of $\Omega$ on $L$ instead of $\ell$, the
latter appearing explicitly in the quark dispersion laws.

In order to avoid confusion with existing literature, we stress that in this paper we shall refer to $\sigma$ as the
order parameter of chiral symmetry breaking and not to the chiral condensate. The identification of the two quantities
is correct in the case of a local interaction (apart an irrelevant multiplicative constant), but in the case of a
non-local interaction they are no longer equal. This has been already discussed in a transparent way in
Ref.~\cite{Hell:2008cc}. The difference arises from the very definition of $\sigma$ in Eq.~\eqref{eq:condensates} in
terms of the {\em dressed} quark fields $q$. This is still more evident by inspection of the gap equation and the self
consistent equation for the chiral condensate. The former is obtained by the requirement that $\Omega$ has to be
minimized by the physical value of $\sigma$. At $T=0$, this leads to the relation
\begin{equation}
\sigma=-2G N_c N_f\int\frac{d^3\bm p}{(2\pi)^3}f(p)^2\frac{M(p)}{\sqrt{p^2 + M(p)^2}}~,
\end{equation}
where we have taken the limit $gH=0$ (for $gH\neq0$ the same reasoning applies) and $L=1$ (that is pure NJL model). On
the other hand, the chiral condensate is defined in terms of the trace of the fermion propagator and of the {\em bare}
quark fields,
\begin{equation}
\langle\bar\psi \psi\rangle = 2\langle\bar{u} u\rangle = -i\int\frac{d^4p}{(2\pi)^4}\text{Tr}S(p)
\end{equation}
where the trace is over color, flavor and Dirac indices. The previous equation leads to
\begin{equation}
\langle\bar{u} u\rangle = -6\int\frac{d^3\bm p}{(2\pi)^3}\frac{M(p)}{\sqrt{p^2 + M(p)^2}}~. \label{eq:UUUtot}
\end{equation}
In order to take into account only the non-perturbative contributions to the expectation value of the operator $\bar u
u$ in the ground state, we have to average over the non perturbative fluctuations only. To this end, the perturbative
contribution to $\langle\bar u u\rangle$ has to be subtracted from the previous equation. Thus the non perturbative
contribution to $\langle\bar{u} u\rangle$, that has to be identified with the true chiral condensate, is given by
\begin{equation}
\langle\bar{u}u\rangle_{NP} = -6\int\frac{d^3\bm p}{(2\pi)^3}\left[\frac{M(p)}{\sqrt{p^2 + M(p)^2}} -\frac{m}{\sqrt{p^2
+ m^2}} \right]~, \label{eq:k}
\end{equation}
which shows that $\sigma\neq 2G\langle\bar{u} u\rangle$. We will make use of Eq.~\eqref{eq:k} in section III.C.

Before closing this section we recall the definition of the dimensionless susceptibility
matrix~\cite{Fukushima:2003fw,Sasaki:2006ww,Abuki:2008nm}. The dimensionless curvature matrix of the free energy around
its global minima $C$ is given by
\begin{equation}
C\equiv\left(%
\begin{array}{cc}
  C_{MM} & C_{M L} \\
  C_{M L} & C_{L L}\\
\end{array}%
\right)~,
\end{equation}
with matrix elements defined as
\begin{eqnarray}
C_{MM} &=& \frac{1}{T\Lambda}\frac{\partial^2\Omega}{\partial M^2}~, \\
C_{L L} &=& \frac{1}{T\Lambda^3}\frac{\partial^2\Omega}{\partial L^2}~,\\
C_{M L} &=& \frac{1}{T\Lambda^2}\frac{\partial^2\Omega}{\partial \Phi \partial M}~,
\end{eqnarray}
and $M\equiv M(p=0)$. The susceptibility matrix $\hat\chi$ is computed as the inverse of the curvature matrix $C$. We
have
\begin{equation}
\hat\chi=\left(%
\begin{array}{cc}
  \chi_{MM} & \chi_{M L} \label{eq:chiMM}\\
  \chi_{M L} & \chi_{LL}  \\
\end{array}%
\right)~.
\end{equation}
Here $\chi_{MM}$ and $\chi_{L L}$ denote respectively the dimensionless susceptibilities of the constituent quark mass
and of the Polyakov loop.

\section{Results}
We now discuss our results. In our model, we have introduced a form factor which mimics asymptotic freedom of QCD. Thus
the hard momenta $p\gg\Lambda$ are naturally cutoff in our theory. However from a numerical point of view we need to
cut the integrations by hand at a certain value of momentum. In order to do this safely we analyze the numerics of the
form factor specified in Eq.~\eqref{eq:f1} and notice that $f(\Lambda) = 1/\sqrt{2}$, $f(2\Lambda) \approx 9.8\times
10^{-4}$ thus $f(2\Lambda)/f(\Lambda)\approx 1\times 10^{-3}$. This means that our interaction is ineffective at
momenta larger than $2\Lambda$. Therefore, it is a good approximation to implement into the numerical integrals the
condition $p<2\Lambda$, which translates into
\begin{equation}
p_z^2 + p_{\pm,s}^2 \leq 4\Lambda^2~
\end{equation}
for red and green quarks. In order to satisfy this condition, we require firstly
\begin{equation}
p_{\pm,s}^2 \leq \Lambda^2\Rightarrow n + \frac{1}{2} \pm \frac{s}{2}\leq
\text{IntegerPart}\left[\frac{4\Lambda^2}{gH}\right] \equiv N_{MAX}~, \label{eq:Nmax}
\end{equation}
where the upper (lower) sign corresponds to the case $s=1$ ($s=-1$). The allowed values for $n$ lie in the range
$\{0,1,\dots,N_{MAX}-1\}$ for $s=1$ and in $\{0,1,\dots,N_{MAX}\}$ for $s=-1$. For $gH>4\Lambda^2$ we have $N_{MAX} =
0$. For each allowed value of $n$ we require secondly
\begin{equation}
p_z^2 \leq gH\left[N_{MAX} -\left(n+\frac{1}{2} + \frac{s}{2}\right)\right]~.\label{eq:PZ}
\end{equation}
In the following, we show results obtained with the choice given in Eq.~\eqref{eq:Nmax}. We have explicitly verified
the stability of our numerical results by doubling the value of the numerical cutoff, thus using a maximum value of $n$
which is four times larger than that in Eq.~\eqref{eq:Nmax}.

\subsection{Chromomagnetic catalysis of chiral condensation}

\begin{figure*}[f]
\begin{center}
\includegraphics[width=10cm]{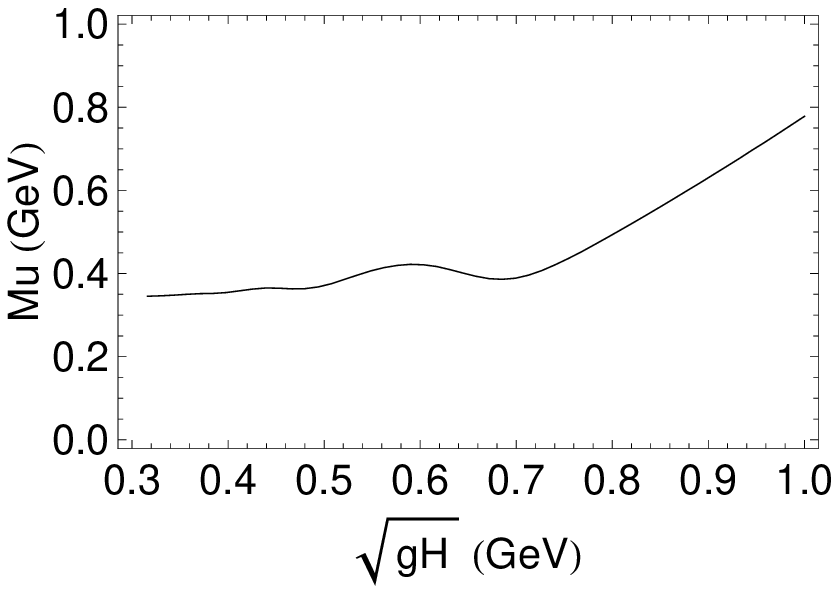}
\caption{\label{Fig:MM} Zero momentum quark mass as a function of the applied chromomagnetic field at zero
temperature.}
\end{center}
\end{figure*}

In Fig.~\ref{Fig:MM}, we plot the zero momentum constituent quark mass $M_u \equiv M(p=0)$ as a function of the
external field at zero temperature. The qualitative behavior of $M_u$ is in agreement with that obtained previously for
NJL in a magnetic field~\cite{Klevansky:1989vi} as well as in a chromomagnetic
field~\cite{Ebert:2000pb,Ebert:2006uh,Klimenko:1993ec} with a proper time regularization. We notice that the net effect
of $gH$ is to increase the value of $M_u$ with respect to its value at $gH=0$. This behavior is the so called
(chromo)magnetic catalysis and it has been interpreted in terms of dimensional
reduction~\cite{Ebert:2000pb,Ebert:2006uh,Gusynin:1995nb,Semenoff:1999xv,Gusynin:1999pq,Miransky:2002rp,Klimenko:1993ec}.

It is interesting to see how dimensional reduction manifests in the model at hand. For $gH/\Lambda^2 \ll 1$ the number
of Landau levels that enter into the gap equation for $\sigma$, namely $N_{MAX}$ defined in Eq.~\eqref{eq:Nmax}, is
extremely large. As a consequence $p_z^2$ can vary in a wide interval, see Eq.~\eqref{eq:PZ}. In this case the motion
is effectively three dimensional. As the magnitude of the external field is increased $N_{MAX}$ decreases. This is not
an artifact of the model but a mere consequence of the existence of a form factor in the interaction that naturally
cuts off the hard momenta $p\gg\Lambda$. As $N_{MAX}$ decreases and eventually reaches the unity, the only possible
value of $p_z^2$ for spin up quarks is zero; regarding spin down quarks $p_z^2$ is zero for $n=1$, while varies in the
range $(0,gH)$ for $n=0$. Thus the motion in the lowest Landau level takes place in a two dimensional momentum space
for spin up quarks, that is in the plane orthogonal to the applied field. The same is true for spin down quarks in the
first excited Landau level. Finally the motion of spin down quarks in the lowest Landau level, with $p^2_\perp=0$,
takes place in an one dimensional momentum space.

\subsection{Deconfinement and chiral transitions}

\begin{figure*}[t]
\begin{center}
\includegraphics[width=7cm]{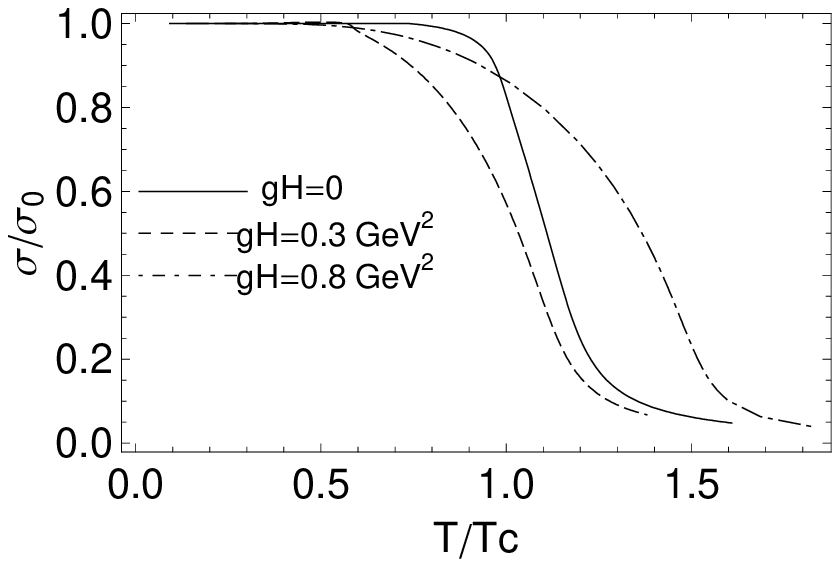}~~~~~\includegraphics[width=7cm]{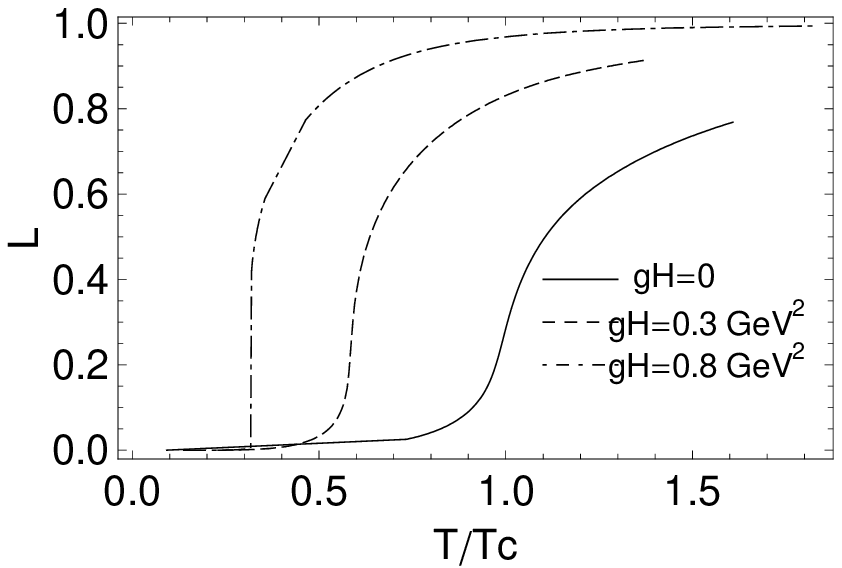}\\
\includegraphics[width=7cm]{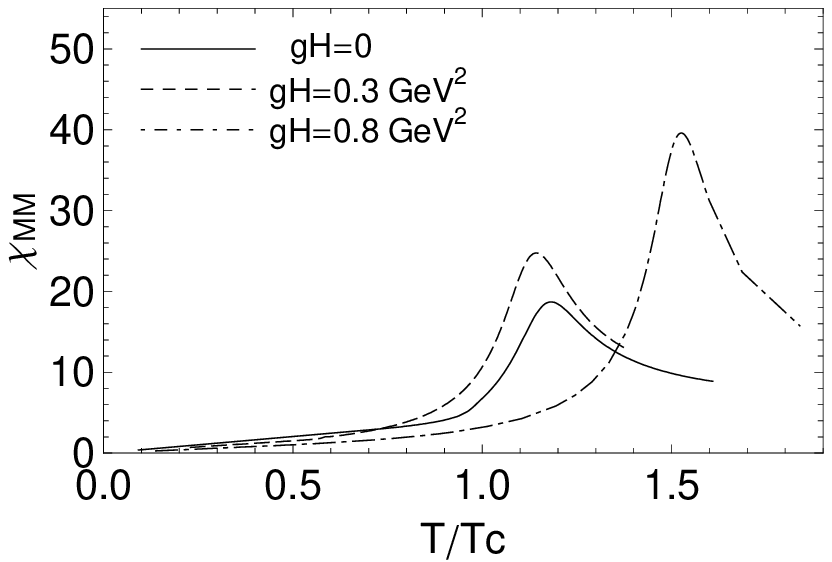}~~~~~\includegraphics[width=7cm]{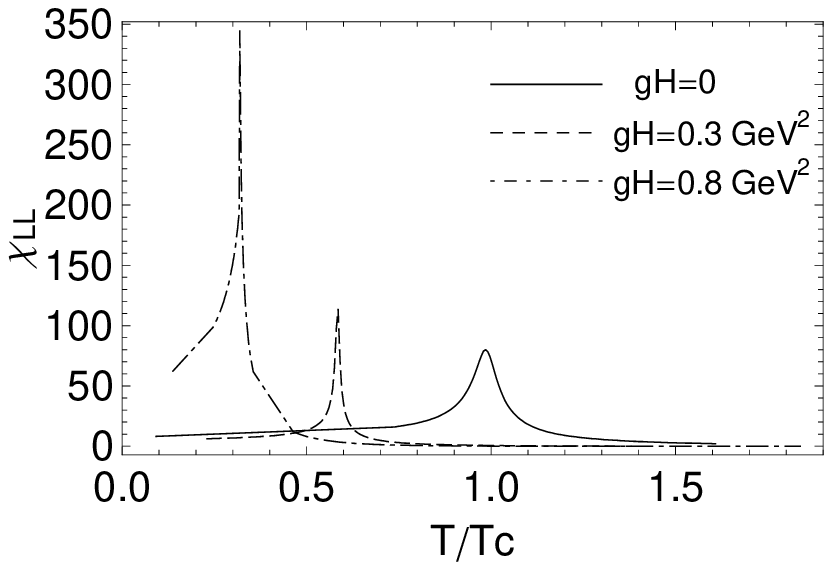}
\caption{\label{Fig:CP} Expectation values of $\sigma$, normalized to its zero temperature value, and of the Polyakov
loop as a function of temperature (in units of the zero field deconfinement temperature $T_c = 217.6$ MeV), for three
different values of $gH$. In the lower panel the chiral and the Polyakov loop susceptibilities are shown.}
\end{center}
\end{figure*}

In Fig.~\ref{Fig:CP}, we plot $\sigma$, normalized for each value of $gH$ to its zero temperature value, the Polyakov
loop and the two physically relevant susceptibilities, $\chi_{MM}$ and $\chi_{LL}$, as a function of temperature for
three values of $gH$. We identify the deconfinement (chiral) crossover with the peak of the Polyakov loop (chiral)
susceptibility $\chi_{LL}$ ($\chi_{MM}$). An interesting feature of the model at hand is that the deconfinememt
crossover becomes a first order transition when the strength of the applied field is larger than a critical value. This
is clear both from the behavior of $\langle L\rangle$ against $T$ and from that of $\chi_{LL}$. As a matter of fact, at
$gH=0$ the Polyakov loop increases smoothly as the temperature is increased. At large values of $gH$ the Polyakov loop
is consistent with zero at small temperature, and becomes suddenly non zero at large temperature. The value of the
critical temperature depends on $gH$. Moreover, $\chi_{LL}$ has a broad and not pronounced peak at $gH=0$, signaling
the transition is actually a smooth crossover (in Fig.~\label{Fig:CP} the value of $\chi_{LL}$ at $gH=0$ is multiplied
by a factor of $10$ for a better comparison with the other data). As $gH$ is increased, $\chi_{LL}$ develops a
pronounced peak, the larger the value of $gH$ the larger the height of the peak. Thus the crossover becomes a true
first order transition.

On the other hand, from the data we obtain for $\chi_{MM}$, we infer that the chiral transition remains a crossover at
each value of $gH$ we have considered. The effect of $gH$ on chiral symmetry breaking is twofold. At zero temperature
we find that the larger the magnitude of $gH$ the larger the magnitude of the chiral condensate. This is in agreement
with the aforementioned scenarios of chromomagnetic catalysis. At finite temperature the scenario changes depending on
the value of $gH$. The chiral crossover temperature slightly decreases at small $gH$, but increases at large $gH$.

\begin{figure*}[bt]
\begin{center}
\includegraphics[width=10cm]{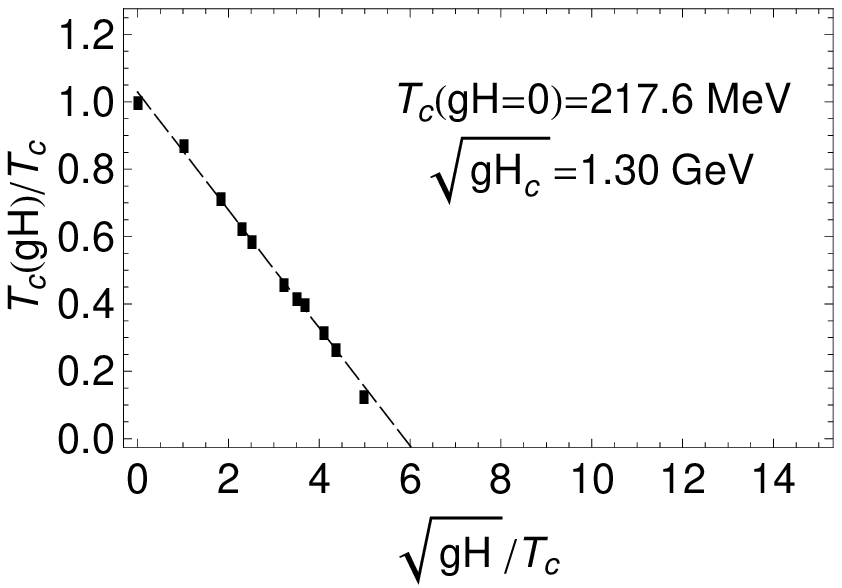}
\caption{\label{Fig:PD} Deconfinement temperature against $\sqrt{gH}$ for the model with dynamical quarks. $T_c$ is
defined as the deconfinement temperature at zero field, $T_c = 217.6$ MeV. The black rectangles correspond to the
results of our calculation; the dashed line is a linear fit to the data. }
\end{center}
\end{figure*}

In Fig.~\ref{Fig:PD}, we plot the critical temperature for deconfinement as a function of the applied field strength
$\sqrt{gH}$.  We identify the deconfinement temperature with the peak of the Polyakov loop susceptibility. In the
figure, the black rectangles are the results of our calculations. Interestingly enough, we can fit the data with a
linear function of $\sqrt{gH}$. We define $T_c$ as the deconfinement temperature at $gH=0$ with dynamical quarks,
namely $T_c=217.6$ MeV. Then we find the best fit to our data as
\begin{equation}
\frac{T_c(gH)}{T_c} = 1-0.176\frac{\sqrt{gH}}{T_c}~,\label{eq:ooo}
\end{equation}
with a linear regression coefficient $R^2=0.996$. Eq.~\eqref{eq:ooo} can be considered the main result of our
investigation. The linear dependence of the deconfinement temperature on the square root of the external field strength
has been noticed for the first time in~\cite{Cea:2002wx,Cea:2005td} within lattice calculations in the pure gauge
theory. Existing Lattice data sustain this picture even in presence of dynamical quarks~\cite{Cea:2007yv}, at least for
$\sqrt{gH}$ of the order of 1 GeV (see for example Fig.~7 of Ref.~\cite{Cea:2007yv}). In our model we use the linear
fit of~\cite{Cea:2002wx,Cea:2005td} on the deconfinement temperature of the {\em pure gauge} as an {\em input}. On the
other hand, the linear dependence of $T_c(gH)$ on $\sqrt{gH}$ with dynamical quarks is an {\em output} of our
calculations. Hence our result strengthens the idea that the PNJL model captures the essential characters of the
deconfinement transition.

It is also interesting to notice that there exists a critical field, that we denote by $\sqrt{gH_c}$, at which the
critical temperature for deconfinement vanishes. This is a mere consequence of the linear dependence depicted in
Eq.~\eqref{eq:ooo}. By requiring $T_c(gH_c) = 0$, we find from Eq.~\eqref{eq:ooo}
\begin{equation}
\sqrt{gH_c}= 1.30~\text{GeV}~,~~~~~\text{with dynamical quarks}~,\label{eq:we}
\end{equation}
in good agreement with the value $\sqrt{gH_c} = 1.6$ GeV obtained by an extrapolation of the Lattice
data~\cite{Cea:2007yv}. The existence of a critical chromomagnetic field for deconfinement is supported by the results
of Ref.~\cite{Ebert:2000pb}: As a matter of fact, it has been shown there that for strong enough chromomagnetic fields
the color superconductivity, thus a deconfined phase, occurs even at zero chemical potential.

\subsection{Changing the form factor in the chiral limit}
In this section, we study the effect of choosing different form factors on the results discussed above. To begin with,
we show that the oscillations of the function $M_u(gH)$ in Fig.~\ref{Fig:MM} is definitely due to the sharpness of the
form factor: choosing a smoother form factor the oscillation disappears. For our purposes, it is enough to take the
chiral limit and to focus on the lorentzian form factors $f_N$ defined by the equation
\begin{equation}
f_N(p)=\frac{1}{\sqrt{1+(p^2/\Lambda^2)^N}}~,~~~~~p=|\bm p|~. \label{eq:fN}
\end{equation}

\begin{figure*}[tb]
\begin{center}
\includegraphics[width=7cm]{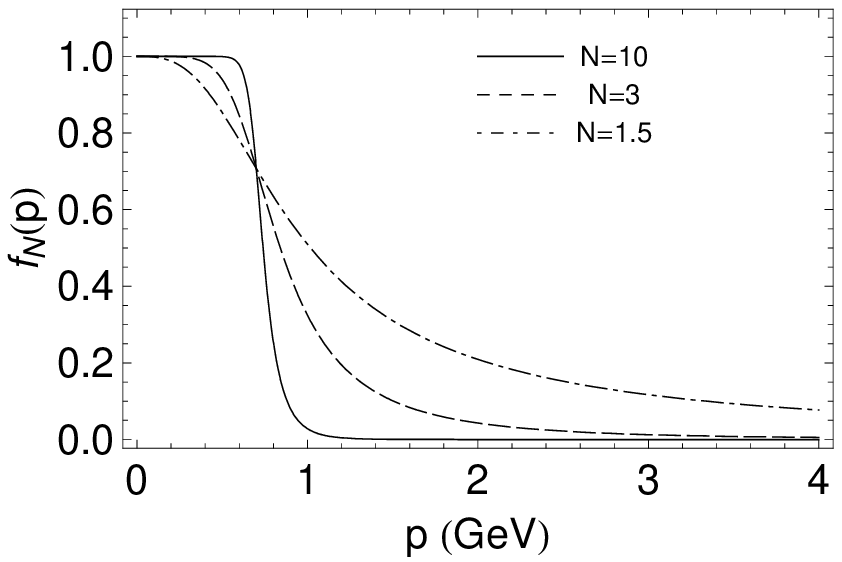}~~~~~\includegraphics[width=7cm]{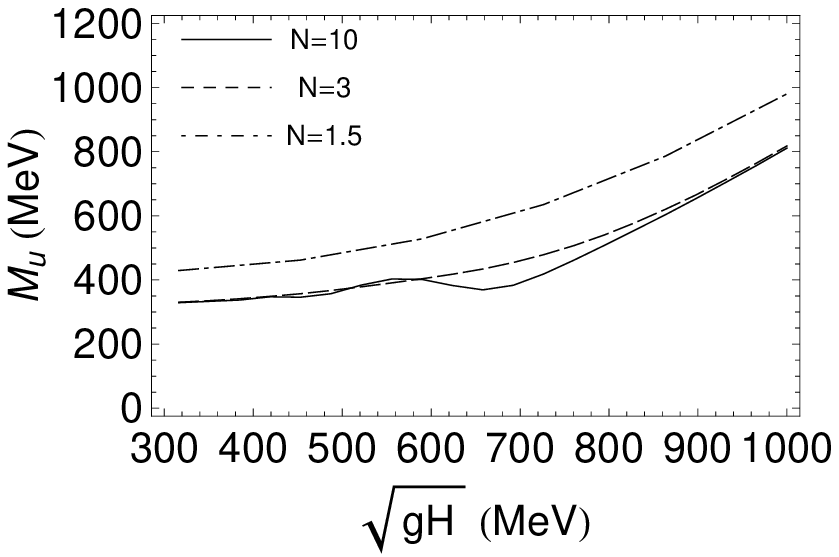}\\
\includegraphics[width=7cm]{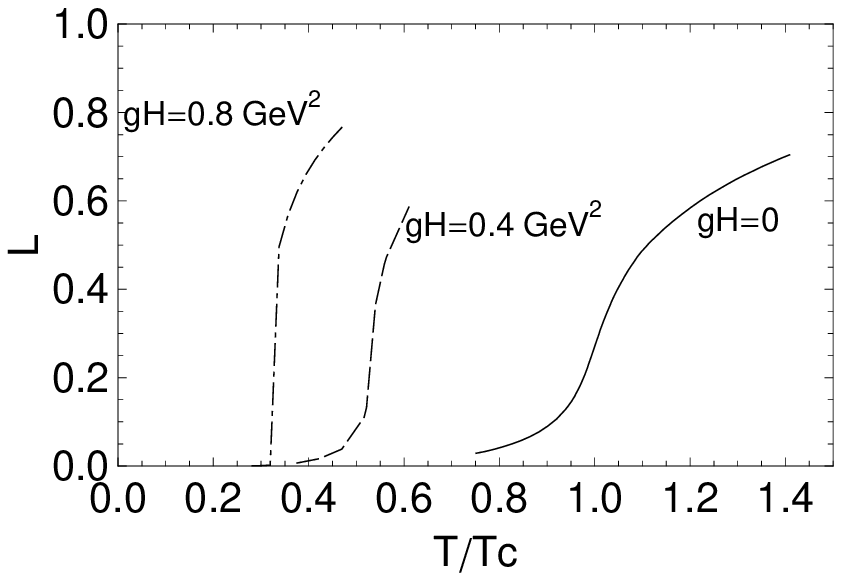}~~~~~\includegraphics[width=7cm]{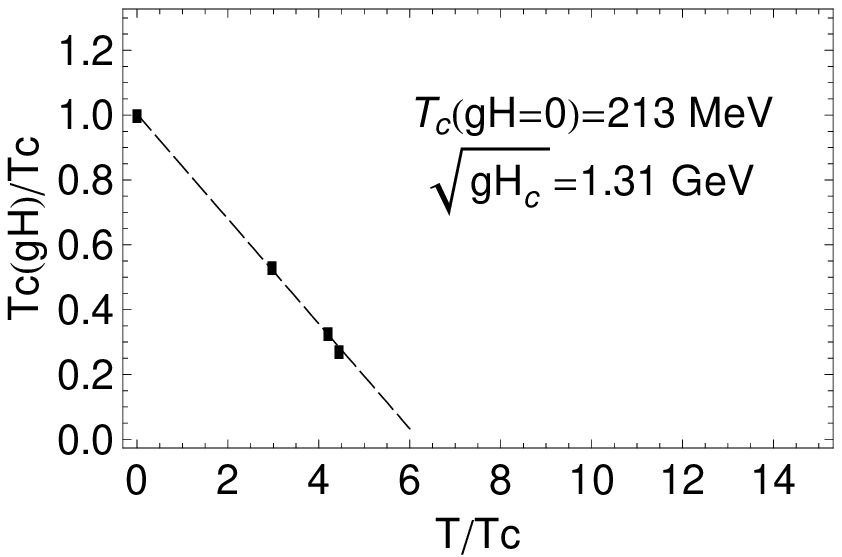}
\caption{\label{Fig:333} Upper left panel: form factors for three different values of $N$ for a common value of
$\Lambda=700$ MeV. Upper right panel: quark self energy at zero $3$-momentum as a function of the applied field
strength. Lower left panel: expectation value of the Polyakov loop computer for three different values of the applied
field strength, in the chiral limit and for the $f_3(p)$ form factor. Lower right panel: deconfinement critical line
computed in the chiral limit and for the $f_3(p)$ form factor. In the lower panels the deconfinement temperature at
zero field is $T_c = 213$ MeV.}
\end{center}
\end{figure*}

In the previous sections, we have shown the results corresponding to $N=10$. We now consider other cases and compare
them to the case $N=10$. As $N$ is decreased, the form factor becomes a smoother function of $p/\Lambda$. This is shown
in Fig.~\ref{Fig:333} where the form factors corresponding to $N=10,3$ and $1.5$ are shown for a common value of
$\Lambda=700$ MeV.

In order to properly fix the parameters $G$ and $\Lambda$ for any given value of $N$ in Eq.~\eqref{eq:fN}, we require
that the model reproduces the experimental value of $f_\pi=92.3$ MeV and a phenomenological value of the constituent
quark mass at zero momentum and at zero field. We take the latter in the range $300-400$ MeV, corresponding to an
expectation value of $|\sigma|$ in the range $150-200$ MeV (for recent estimates of this quantity see for
example~\cite{Hell:2008cc,Bowman:2002kn}). Once the parameters are fixed at a given value of $N$, we measure the value
of the subtracted chiral condensate defined by Eq.~\eqref{eq:k}, taking care it is in the allowed phenomenological
range $190$ MeV $\leq$ $-\langle\bar{u}u\rangle_{NP}^{1/3}$ $\leq$ $260$ MeV~\cite{Dosch:1997wb}. This constraint on
$\langle\bar{u}u\rangle_{NP}^{1/3}$ can be fulfilled only for relatively large values of $N$, namely $N\geq5$, within
the instantaneous form factor approach~\cite{Grigorian:2006qe,GomezDumm:2006vz}; on the other hand, more refined
calculations based on non-instantaneous form factors can fulfill the above
constraint~\cite{Hell:2008cc,GomezDumm:2006vz} also for smoother form factors. Therefore the picture we draw in this
section has to be considered purely qualitative, at least for relatively small values of the parameter $N$. The
numerical values of the parameters are collected in Table~\ref{Tab:para}.

\begin{table}
\begin{tabular}{|c|c|c|c|c|}
\hline\hline
 & $\sigma$~(MeV) &$\Lambda$ (MeV)& $G\Lambda^2$ & $-\langle\bar{u}u\rangle_{NP}^{1/3}$ \\
 \hline
$f_{10}$ & -160.2&660& 2.36& 256 \\
 \hline
$f_{5}$ & -160.2&679& 2.49& 269 \\
 \hline
$f_{3}$ & -170.2&674& 2.62& 282 \\
 \hline
$f_{3/2}$ & -200&614& 2.51& 312 \\
 \hline
\end{tabular}
\caption{\label{Tab:para} Parameters for the various form factors examined. For all the values of $N$ the inputs are
$f_\pi=92.3$ MeV and the value of $\sigma$; $\Lambda$, $G\Lambda^2$ and $-\langle\bar{u}u\rangle_{NP}^{1/3}$ are
outputs.}
\end{table}

In the upper right panel of Fig.~\ref{Fig:333}, we plot the quark self energy at zero momentum as a function of the
applied field strength. We have shown results for $N=10$ (solid line), $N=3$ (dashed line) and $N=3/2$ (dot-dashed
line). As anticipated the effect of a smoother form factor is to reduce (and eventually definitively damp) the
oscillations observed for the case $N=10$.

We have checked the stability of our main results on the deconfinement temperature in external field by changing the
form factor to a smoother one in the chiral limit. The results are summarized in the lower panels of
Fig.~\ref{Fig:333}, where we plot the expectation value of the Polyakov loop computed for the form factor $f_3(p)$, and
the deconfinement line. In this case the deconfinement temperature at zero field is $T_c = 213$ MeV. The dashed line
corresponds to the linear fit to the data, specified by the equation
\begin{equation}
\frac{T_c(gH)}{T_c} = 1-0.161\frac{\sqrt{gH}}{T_c}~,\label{eq:ooo3}
\end{equation}
with a linear regression coefficient $R^2=0.99$. In this case, the critical field corresponding to a vanishing
deconfinement temperature is
\begin{equation}
\sqrt{gH_c}= 1.31~\text{GeV}~,
\end{equation}
in agreement with our previous estimate obtained using $N=10$ and with quarks in the physical limit, see
Eq.~\eqref{eq:we}.

\subsection{The phase diagram in the $gH-T$ plane}

\begin{figure*}[tb]
\begin{center}
\includegraphics[width=10cm]{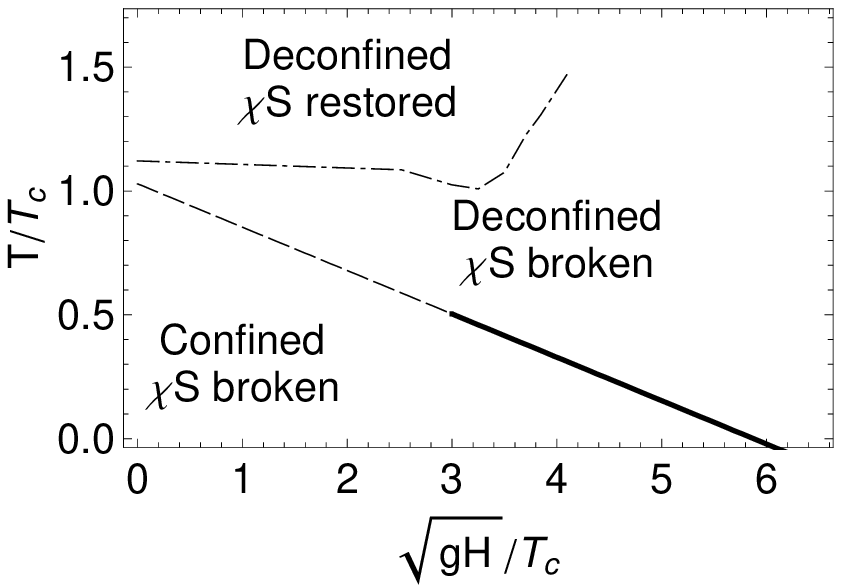}
\caption{\label{Fig:PD3} Phase diagram of the PNJL model in the $gH-T$ plane. Mass scale is given by $T_c = 217.6$ MeV
corresponding to the deconfinement temperature of the model with dynamical quarks and at $gH=0$. Dashed line
corresponds to the deconfinement crossover; solid line denotes the first order deconfinement transition. Dot-dashed
line corresponds to the chiral crosser.}
\end{center}
\end{figure*}

We summarize our results in a phase diagram in the $gH-T$ plane, see Fig~\ref{Fig:PD3}. In the figure, the dashed line
denotes the deconfinement crossover, which becomes a first order transition as $gH$ is larger than a critical value,
see the bold line in Fig.~\ref{Fig:PD3}. Both the dashed and the solid lines are the results of the fit given in
Eq.~\eqref{eq:ooo} and are identified by the position of the peaks of the Polyakov loop susceptibility.  They are an
output of our calculations once the input Eq.~\eqref{eq:T0mM} is given. Our results show that there exists a critical
endpoint in the phase diagram, with coordinates given by
\begin{equation}
(\frac{\sqrt{gH_E}}{T_c},\frac{T_E}{T_c}) = (3.1,0.5)~,\label{eq:endP}
\end{equation}
that is,
\begin{equation}
(\sqrt{gH_E},T_E) = (0.653,0.108)~\text{GeV}~.\label{eq:endP}
\end{equation}
The critical end point coordinates have been computed by the investigation of the Polyakov loop temperature dependence.
At a given value of $gH$, we compute $L$ as a function of temperature. If the function $L(T)$ is continuous then we
identify a crossover. On the other hand, if $L(T)$ shows a discontinuity, then we identify the critical temperature by
the temperature at which the discontinuity occurs, and identify the transition with a first order transition. By
iteration of this process for several values of $gH$, we are able to determine the critical end point coordinates. The
uncertainty of our final result depends only on the size of the interval of $\sqrt{gH}$, around $gH_E$, in which we
look for $L(T)$, and we can estimate it to be of the order of 10 MeV, both on $T$ and on $\sqrt{gH}$

In Fig.~\ref{Fig:PD3} we draw also the chiral crossover line. It is obtained by the position of the peaks in the chiral
susceptibility. For $gH<gH_E$, the effect of the applied field is to slightly lower the transition temperature. For
$gH>gH_E$, the slope of the chiral crossover line changes sign thus opening a large window with deconfined matter but
with chiral symmetry spontaneously broken. The existence of the valley in the chiral critical line around $gH_E$ could
be an artifact of the specific functional choice of the form factor. On the other hand, the qualitative behavior we
have depicted should be quite robust since the dimensional reduction and the consequent chromomagnetic catalysis exist
in the model at hand (see the discussion at the beginning of this section) independently on which form factor is
chosen.

\subsection{Equation of state in the external field}
For completeness, we compute in this section the equation of state $p=p(\varepsilon)$ of the PNJL model for different
values of the strength of the external field. This kind of computation is interesting because the equation of state is
accessible to Lattice measurements, hence our results can be directly checked. The various thermodynamical quantities
are defined as follow:
\begin{eqnarray}
p &=& -\Omega~, \\
s &=& -\frac{\partial\Omega}{\partial T}~,\\
\varepsilon &=& -p + Ts~,\\
C_V &=& -T\frac{\partial^2\Omega}{\partial T^2}~,\\
c_s^2 &=& \frac{dp}{d\varepsilon} = \frac{s}{C_V}~.
\end{eqnarray}
In the above equations $p$, $s$, $\varepsilon$, $C_V$ and $c_s^2$ denote the pressure, the entropy, the internal energy
density, the specific heat and the squared sound velocity, respectively. We normalize the pressure with the convention
$p(T=0)=0$ for each value of $gH$.

\begin{figure*}[b]
\begin{center}
\includegraphics[width=8cm]{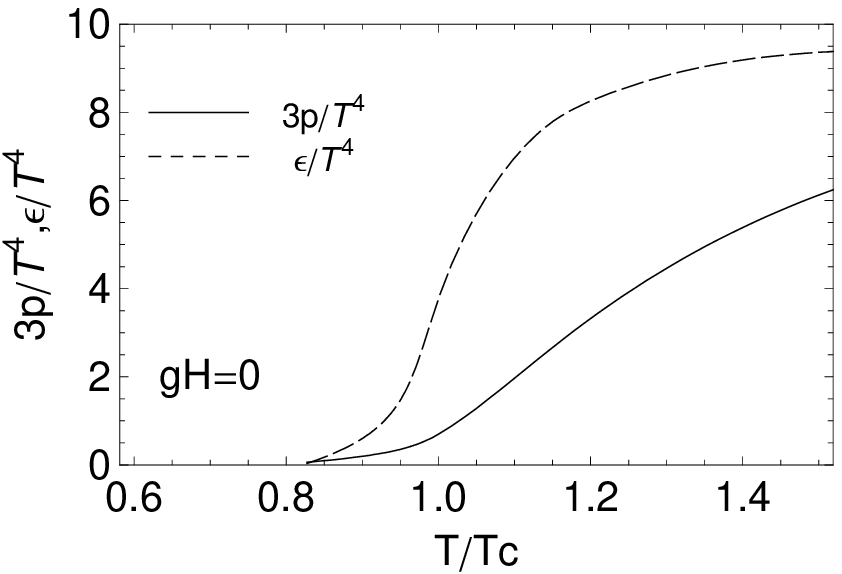}~~~~~\includegraphics[width=8cm]{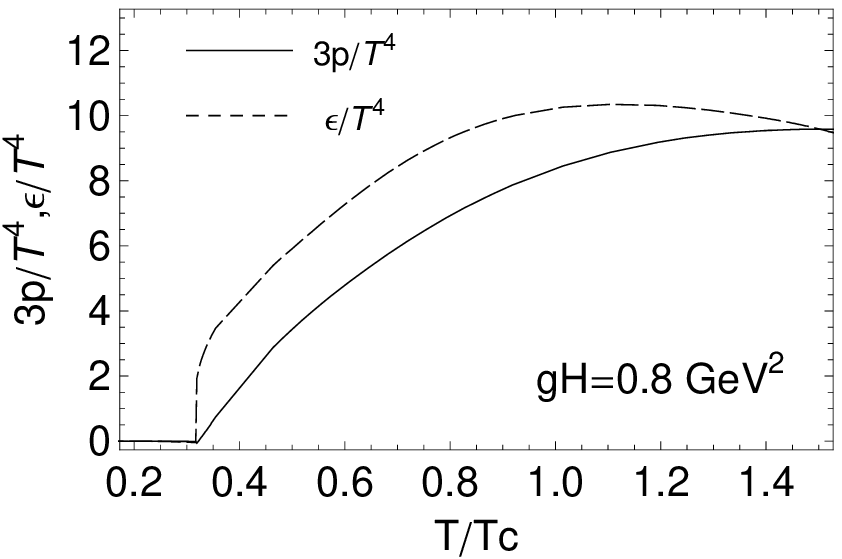}\\
\includegraphics[width=8cm]{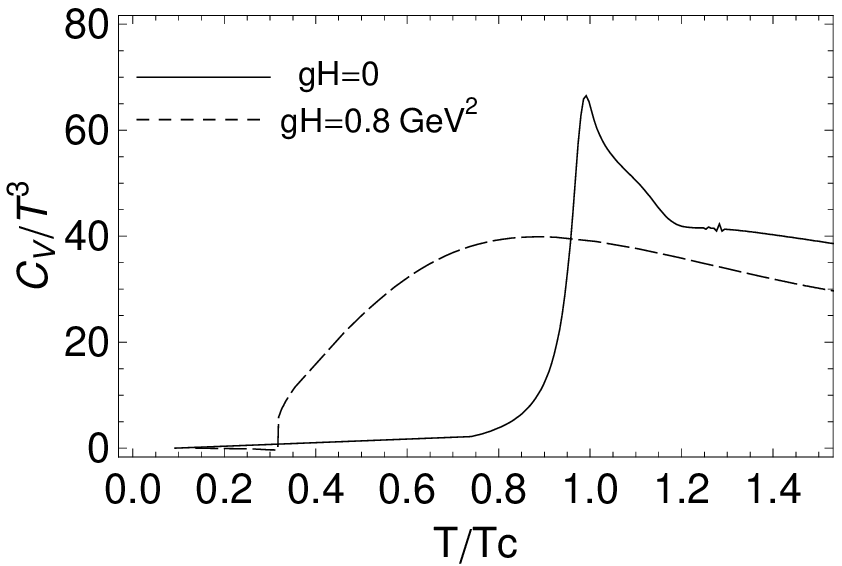}~~~~~\includegraphics[width=8cm]{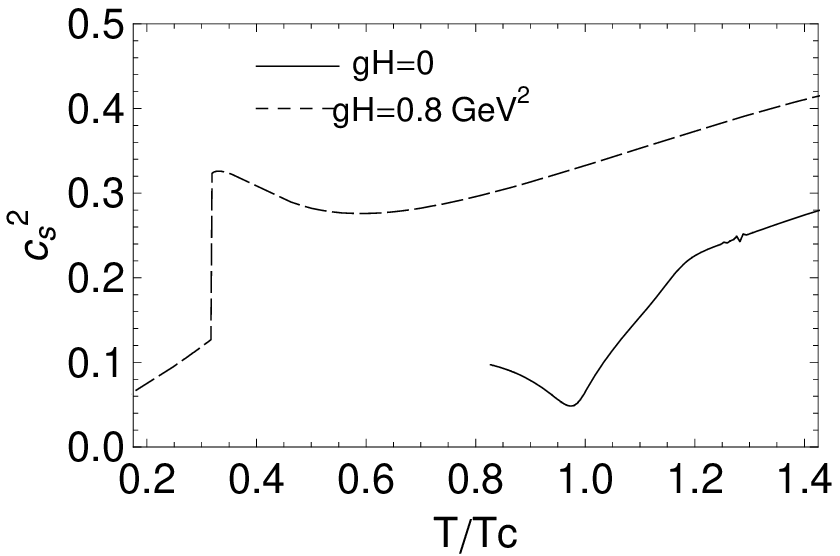}
\caption{\label{Fig:5} Upper left panel: pressure and energy density, normalized to $T^4$, against temperature in the
case $gH=0$. Upper right panel: pressure and energy density, normalized to $T^4$, against temperature in the case
$gH=0.8$ GeV$^2$. Lower left panel: dimensionless specific heat $C_V/T^3$ as a function of temperature for $gH=0$ and
$gH=0.8$ GeV$^2$. Lower right panel: squared sound velocity against temperature. In the figure the deconfinement
temperature at zero field is $T_c = 217.6$ MeV.}
\end{center}
\end{figure*}

\begin{figure*}[b]
\begin{center}
\includegraphics[width=8cm]{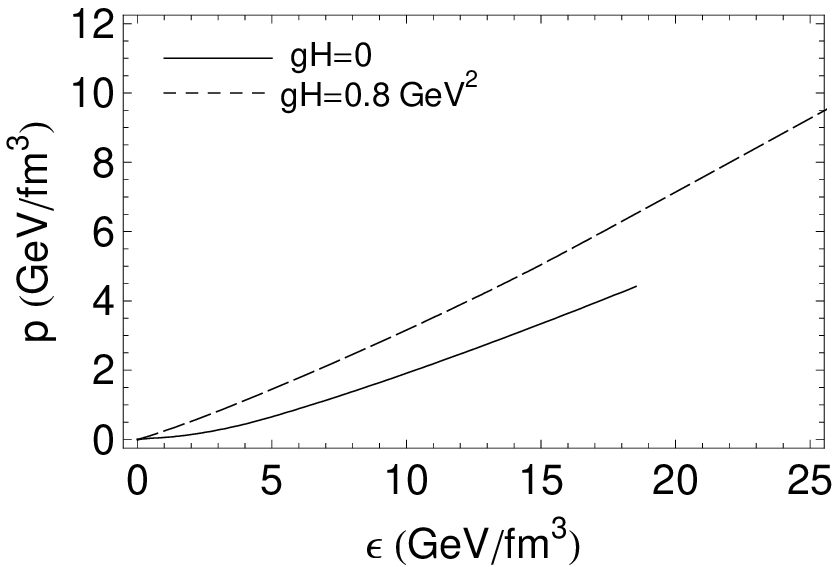}~~~~~\includegraphics[width=8cm]{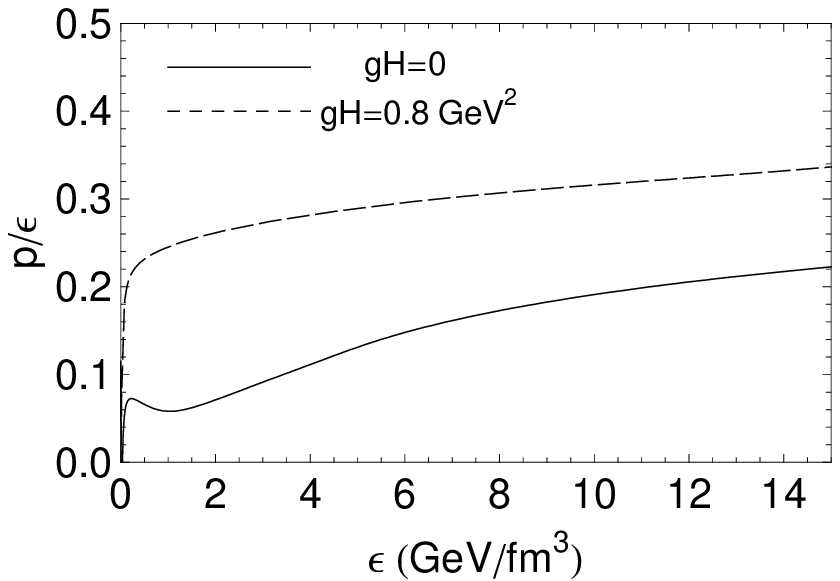}
\caption{\label{Fig:7} Equation of state of PNJL matter at zero field (solid line) and at $gH=0.8$ GeV$^2$ (dashed
line).}
\end{center}
\end{figure*}

The results of our numerical calculations are shown in Figs.~\ref{Fig:5} and~\ref{Fig:7} where we plot the pressure and
the energy density, the specific heat, the squared sound velocity and the equation of state for $gH=0$ and for $gH=0.8$
GeV$^2$. From the qualitative point of view the thermodynamical quantities behave like the $gH=0$ case for $gH<gH_E$,
and like in the case $gH=0.8$ GeV$^2$ for $gH>gH_E$. For this reason, it suffices to plot the various quantities only
for these two values of the applied field strength.

\section{Conclusions}
In this paper, we have investigated the response of the PNJL vacuum to an external abelian chromomagnetic field. Our
interaction is non-local but instantaneous, with a momentum-dependent Lorentzian form factor. We have used as an input
the Lattice-deduced dependence of the deconfinement temperature of pure gauge theory on the strength of the applied
field $gH$, and we have computed the effect of dynamical quarks on deconfinement.

We can summarize our results as follows: An external abelian chromomagnetic field inhibits color confinement even the
in presence of dynamical quarks. Moreover the dependence of the deconfinement temperature on $\sqrt{gH}$ is linear. We
have compared our results with existing Lattice data, the latter exploring values $gH$ up to the order 1 GeV$^2$ in the
case of full QCD, finding qualitative agreement. An extrapolation of our data at larger values of $gH$ leads to the
existence of a critical field $gH_c$ above which quark matter is deconfined even at small temperatures. We have checked
the stability of this result by changing the analytical form of the momentum-space form factor. We have drawn the phase
diagram of the model in the $gH-T$ plane in the physical limit. Finally we have computed the equation of state of the
PNJL matter in the external field.

The results on the deconfinement temperature are encouraging: They suggest that the PNJL model captures the essential
physics of deconfinement. On the other hand we find that the external field favors spontaneous chiral symmetry
breaking, as in the NJL model (chromomagnetic catalysis). The latter result is in disagreement with Lattice, where a
single transition temperature is observed at which both the Polyakov loop and the chiral susceptibilities show
pronounced peaks. Thus, Lattice supports a scenario in which deconfinement and chiral restoration occur at the same
temperature for any value of $gH$; as a consequence spontaneous chiral symmetry breaking should be inhibited by the
external field and not catalyzed. Probably, the discrepancy arises mainly from the fact that our interaction is
instantaneous, that is it depends only on the three-momentum.

A comparison with QCD in strong magnetic fields can be helpful. The authors of Ref.~\cite{Miransky:2002rp} have shown,
within an improved ladder approximation, that in QCD a strong magnetic field, $B$, leads to a dynamical mass $m_q(B)$
which is lower than $m_q(B=0)$, if $eB$ is in the range $\Lambda_{QCD}^2 \lesssim eB \lesssim (10~\text{TeV})^2$. This
is mainly due to the running of $\alpha_s$ at the scale $eB$,
\begin{displaymath}
\frac{1}{\alpha_s} \propto \log\frac{|eB|}{\Lambda_{QCD}^2}~.
\end{displaymath}
Moreover, in the momentum region relevant for dynamical symmetry breaking, that is $m_q^2\lesssim k_0^2-\bm k^2
\lesssim eB$, the magnetic field acts as a cutoff in the whole 4-momentum space; on the other hand we have assumed that
all the energy region is relevant for dynamics (we have no cutoff on energy), and our coupling constant $G$ is fixed,
once for all, at the value corresponding to zero field. These two effects combined together probably lead to an
overestimate of the chiral condensate, as argued in Ref.~\cite{Miransky:2002rp}. Investigations based on
non-instantaneous interaction as well as on running coupling are now under investigation and will be discussed in a
forthcoming paper.

\acknowledgments We acknowledge L. Cosmai, P. Cea, R. Gatto, K. Klimenko and M. Mannarelli for interesting discussions
and useful comments.

\end{document}